\documentclass[conference,letterpaper]{IEEEtran}

\pdfoutput=1
\usepackage[utf8]{inputenc} 
\usepackage[T1]{fontenc}
\usepackage{url}
\usepackage{color}
\usepackage{subfigure}
\usepackage{graphicx}
\usepackage{hyperref}
\usepackage{ifthen}
\usepackage{cite}
\usepackage[cmex10]{amsmath} 

\begin{document}
\title{ Gaussian Boson Sampling to Accelerate
NP-Complete Vertex-Minor Graph Classification} 


\author{%
  \IEEEauthorblockN{Mushkan Sureka}
  \IEEEauthorblockA{University of Arizona,
                    Tucson, AZ, USA\\
                    Email: mushkansureka@arizona.edu}
  \and
  \IEEEauthorblockN{Saikat Guha}
  \IEEEauthorblockA{University of Arizona,
                    Tucson, AZ, USA\\
                    Email: saikat@arizona.edu}
}


\maketitle


\begin{abstract}
   Gaussian Boson Sampling (GBS) generates random samples of photon-click patterns from a class of probability distributions that are hard for a classical computer to sample from. Despite heroic demonstrations of quantum supremacy using GBS, Boson Sampling, and instantaneous quantum polynomial (IQP) algorithms, systematic evaluations of the power of these quantum-enhanced random samplers when applied to provably hard problems, and performance comparisons with best-known classical algorithms have been lacking. We propose a hybrid quantum-classical algorithm using the GBS for the NP-complete problem of determining if two graphs are a vertex minor of one another. The graphs are encoded in GBS and the generated random samples serve as feature vectors in a support vector machine (SVM) classifier. We find a graph embedding that allows trading between the one-shot classification accuracy and the amount of input squeezing, a hard-to-produce quantum resource, followed by repeated trials and a majority vote to reach an overall desired accuracy. We introduce a new classical algorithm based on graph spectra, which we show outperforms various well-known graph-similarity algorithms. We compare the performance of our algorithm with this classical algorithm and analyze their time versus problem-size scaling, to yield a desired classification accuracy. Our simulation results suggest that with a near-term realizable GBS device---$5$ dB pulsed squeezers, $12$-mode unitary, and reasonable assumptions on coupling efficiency, on-chip losses, and detection efficiency of photon number resolving detectors---we can solve $12$-node vertex-minor instances with about $10^3$ fold lower time compared to a powerful desktop computer.
\end{abstract}

\section{Introduction}

In recent years, Noisy Intermediate-Scale Quantum (NISQ) processors have been used to generate random samples from complex probability distributions that are hard to sample from using a classical computer. Some examples include: Google's random superconducting circuit~\cite{arute2019quantum}, Xanadu's Gaussian Boson Sampling (GBS)~\cite{madsen2022quantum}, and USTC's Boson Sampling~\cite{Wang2019} demonstrations. In GBS, $N$ independent squeezed-light pulses are sent through an $N$-port passive unitary---a linear-optical interferometer, and each output mode is detected by a photon number resolving (PNR) detector. The click pattern that appears at the $N$ PNR detectors are random samples from a distribution that is classically hard to sample from~\cite{hamilton2017gaussian, Deshpande2022}. Researchers from Xanadu proposed multiple practical applications of GBS, which include: graph similarity~\cite{schuld2019quantum}, point processes~\cite{jahangiri2020point}, graphs problems such as the Maximum Clique~\cite{banchi2020molecular} and the Dense Subgraph problems~\cite{arrazola2018using} and various machine-learning inspired primitives~\cite{schuld2019quantum, jahangiri2020point, banchi2020training}.
\\

  \begin{figure}[t]
    \centering
    \includegraphics[width=0.99\linewidth]{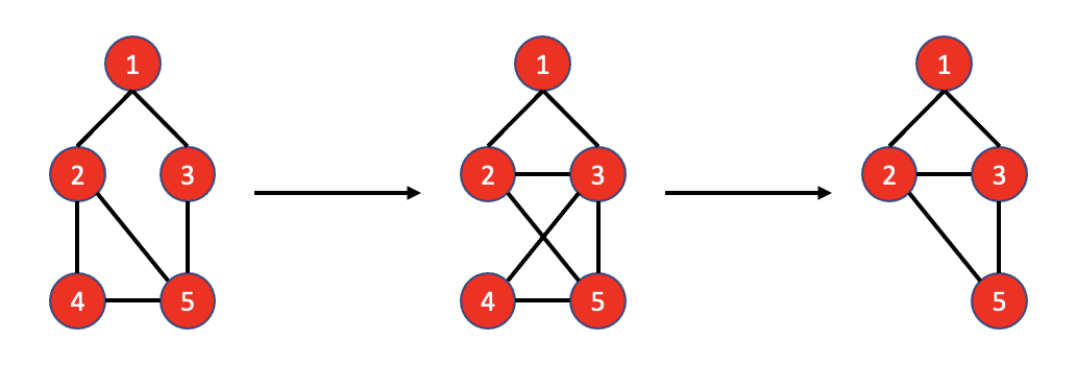}
    \caption{Pictorial representation of obtaining a vertex minor of a graph. The 1$^{\rm st}$ graph undergoes local complementation at vertex 5, which results in the 2$^{\rm nd}$ graph. The 3$^{\rm rd}$ graph is obtained by vertex deletion of node 4. The third graph is called a vertex minor of the first graph, i.e., one that is obtained via local complementations applied at a collection of vertices and a deletion of a collection of vertices, on the original graph.}
    \label{fig:vm}
\end{figure}
In this paper, we consider a GBS-based solution to the NP-Complete decision problem of determining if one graph is a vertex minor of another~\cite{dahlberg2022complexity}. A graph $G^\prime$ is a vertex minor of graph $G$ if $G^\prime$ can be obtained via a series of {\em local complementations} (LCs)~\cite{Adcock2020} and {\em vertex deletions} starting with $G$ (see Fig.~\ref{fig:vm} for an illustration of a 4-node graph that is a vertex minor of another 5-node graph). Our main results are:
\begin{enumerate}
  \item We propose a randomized classical algorithm that utilizes LCs and graph spectra in a Support Vector Machine (SVM) to decide whether one is a Vertex Minor of another. We show this outperforms several well-known classical algorithms for graph classification. 
  \item We propose a hybrid quantum-classical algorithm that encodes the two graphs in two GBS devices and uses the photon-click-pattern samples as feature vectors in an SVM classifier. We modify the graph embedding algorithm from Ref.~\cite{schuld2019quantum}, allowing us to trade between the one-shot sampling accuracy (i.e., from one run of the GBS) with the maximum amount of per-mode squeezing used at the input. The overall desired classification accuracy is attained by several runs of the GBS and a majority vote decision at the end to declare the final verdict. We argue a runtime versus problem size complexity for our algorithm, which lends evidence towards its time-scaling superiority over using a classical computer (even though both ultimately will have an exponential time complexity, given the problem is NP-complete).
  \item We show that using (near-term realizable) $5$ dB pulsed squeezing per mode at the input of a $12$-mode on-chip GBS generated using a nonlinear spontaneous four-wave mixing (sFWM) process with a mode-locked fibre laser pump of $10$ MHz repetition rate, 80$\%$ coupling efficiency (squeezing chip to photonic integrated circuit (PIC) realizing the interferometer), $0.25$ dB/cm on-chip losses, and 95$\%$ PNR detector efficiency, with an appropriate number of repeated trials (to get the final classification accuracy up to a desired value), our algorithm outperforms the abovesaid classical algorithm (also provided with an appropriate number of repeated trials to get to the same desired accuracy)---run on a MacBook Pro (hardware specifications in Appendix:\ref{appendix:features})---for up to $12$ node vertex-minor problems, by anywhere between a factor of $10^2$ to a factor of $10^3$ in runtime.
\end{enumerate}

\section{Background}
\subsection{Encoding a graph in the GBS}
 
Consider a graph $G(V,E)$ defined over a vertex set $V$ and edge set $E$, with $|V| = N$ nodes. Let us say its adjacency matrix is $A$, which is a $N \times N$ symmetric matrix. Define:
 \begin{equation}\label{eq:1}
 \Tilde{A}= c \begin{pmatrix}
  A & 0\\ 
  0 & A
\end{pmatrix} = c (A\oplus A),
 \end{equation}
 where $0 < c < 1/s_{\rm{max}}$ is a constant and $s_{\rm{max}}$ is the maximum singular value of $A$. Let us consider an $N$-mode pure Gaussian state defined uniquely by a $2N \times 2N$ covariance matrix $\sigma$, with
 \begin{equation}
  \label{eq:2}
\sigma= Q- I/2 \textnormal{ with } Q= (I-X\Tilde{A})^{-1}, X= \begin{pmatrix}
  0 & I\\ 
  I & 0
\end{pmatrix},
 \end{equation}
where $I$ is the $2N \times 2N$ identity matrix. If we perform photon number resolving (PNR) detection on each of the $N$ modes of this state, we would get a random vector of photon number outcomes (henceforth called a GBS {\em sample}) $\textbf{n}$= [$n_1,....,n_M$], where $n_i$ is the number of photons detected in the $i$-th mode, the probability mass function (p.m.f.) of which is given by~\cite{schuld2019quantum}:
\begin{equation} \label{eq:3}
    p(\textbf{n})= \dfrac{1}{\sqrt{\textnormal{det}Q}\, \textbf{n}!} \textnormal{Haf}^2(A_{\textbf{n}}),
\end{equation}
where $\textbf{n}! = n_1 ! n_2 ! \ldots n_N !$, and $A_{\textbf{n}}$ is a matrix that contains $n_j$ duplicates of $j$-th rows and columns of $A$. Appendix \ref{appendix:encode} outlines how choice of $c$ affects the input squeezing per mode.
\subsection{Vertex Minor}
 \begin{figure}
    \centering
    \includegraphics[width=0.99\linewidth]{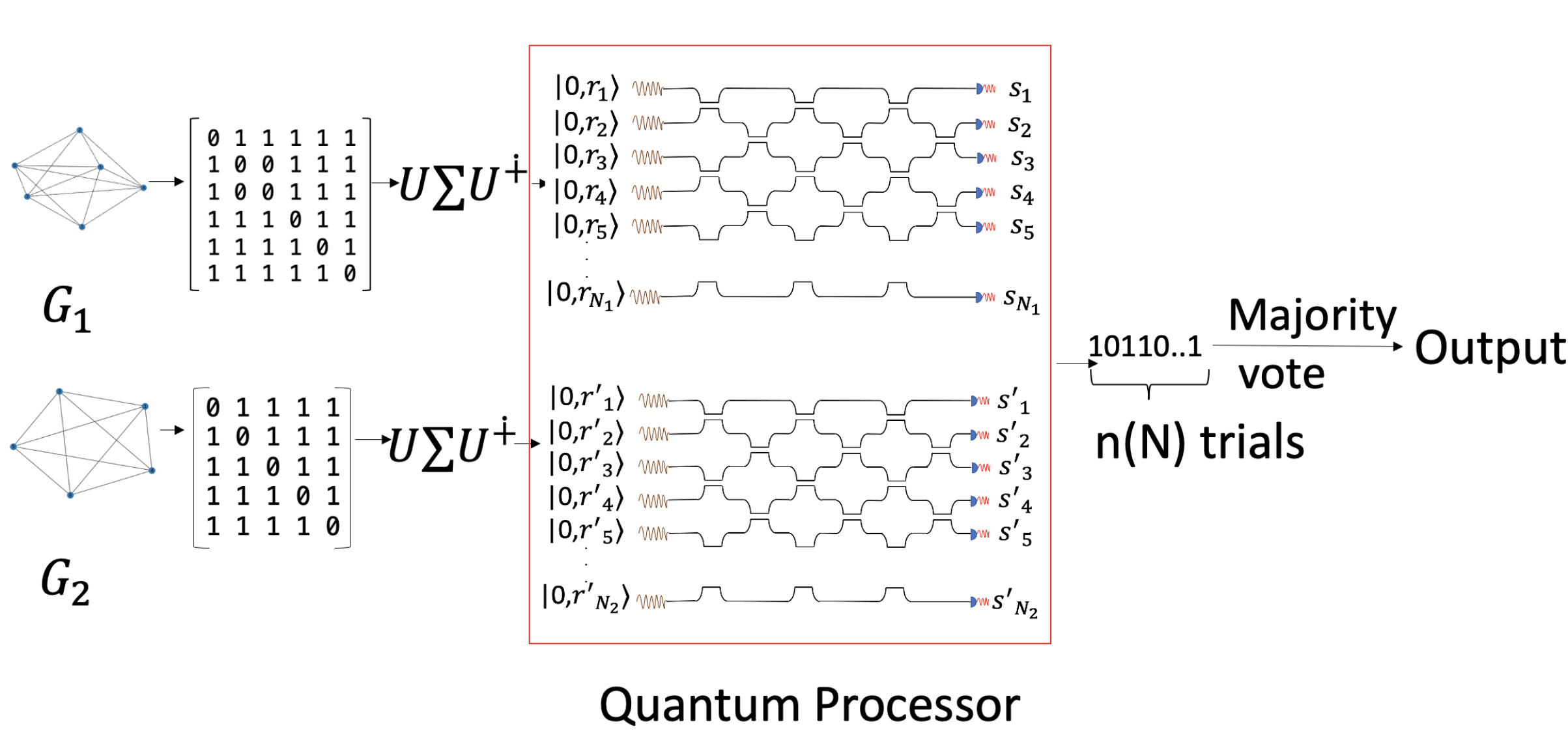 }
    \caption{A conceptual framework for the implementation of our quantum-accelerated algorithm for the vertex minor problem. The algorithm takes two graphs, denoted as $G_1$ and $G_2$, as inputs. Samples are extracted from the GBS programmed using Autonne-Takagi decomposition values of the adjacency matrices. The sampling algorithm is executed multiple times (and results are aggregated using majority voting) until desired overall accuracy is met.
}
    \label{fig:structure}
\end{figure}
LC on a vertex $v$ of a graph $G$ results in a graph $\tau_v(G)$ obtained by inverting the edges between all vertex pairs in the neighbourhood of $v$ in $G$. Deletion of a vertex $v$ of graph $G$ (and all its adjacent edges) results in a graph $\chi_v(G)$. A graph $G^\prime$ is a vertex minor of a graph $G$ if there exists a sequence of LCs and vertex deletions, applied to $G$, which yields $G'$. A vertex minor $G^\prime$ of $G$ can be obtained by having vertices ($v_1,....,v_j$) of $G$ undergo local complementations, i.e., $H = {\tau_{v_j}}\circ{{\tau}_{v_{j-1}}} \ldots \circ{\tau_{v_1}}(G)$, followed by vertex deletions on vertices ($u_1,....,u_k$) of $H$, leading to: $G^\prime = \chi_{u_k}\circ \chi_{u_{k-1}} \ldots \circ \chi_{u_1}(H)$, where $k$ and $j$ are arbitrary integers. 

Both of the above two graph operations---when graph $G$ is associated with a quantum {\em graph state}---correspond to local single-qubit operations~\cite{Raussendorf2001}. LC is a single-qubit Clifford unitary, and vertex deletion can be realized by a Pauli $Z$ measurement on a vertex. Therefore, a sequence of LCs and vertex deletions correspond to a sequence of single-qubit Clifford unitaries and measurements~\cite{dahlberg2020transform}. The decision problem to determine if $G^\prime$ is a vertex minor of $G$ is an important problem underlying Clifford manipulation of Stabilizer quantum states, but is known to be NP-Complete~\cite{dahlberg2022complexity}. This is our primary motivation behind the study of this problem as a candidate for quantum acceleration by a NISQ processor.

\subsection{Classical algorithms for the vertex minor problem}
We explored and implemented a few classical algorithms from the graph similarity literature, and compared their accuracies and runtimes. Brief synopses of these algorithms are below, with further details in Appendix~\ref{appendix:classical}.

\begin{enumerate}
  \item\emph{Graphlet kernel}---exploits a similarity measure for graphs that captures local structural information. It computes the frequency of graphlets (small subgraph) within both graphs and compares their distributions. 
  \item\emph{Shortest path kernel}---is based on a graph kernel that measures the similarity between two graphs based on their shortest-path distances. 
  \item\emph{Weisfeiler-Lehman Kernel}---is a graph kernel that compares the structural similarity of two graphs by counting the number of common substructures in their Weisfeiler-Lehman subtree patterns. It iteratively labels the nodes with the local patterns of their neighbours and aggregates these labels to form a global representation of the graph. 
  \item\emph{Image classification on ordered adjacency matrices}---is a method of classifying graphs wherein their adjacency matrices are converted into images after conditioning them, which are then fed into a neural network.
  \item\emph{Spectral method}---is a simple algorithm we propose and evaluate, which does surprisingly well (and much better than all the above four algorithms) for small instances of the vertex minor problem. We concatenate eigenvalues of Laplacians of the two graphs to construct a feature vector, which is used as the feature in an SVM classifier. 

\end{enumerate}

\section{Methods}

\subsection{Brute-force method to generate training data}

To train the SVM backend, both for our classical (spectral) method as well as the quantum (GBS) method, we generated a data set comprising $500$ vertex-minor graph pairs, and $500$ non-vertex-minor graph pairs, by brute force. We pick two graphs $G$ and $G^\prime$. Let us assume $N > N^\prime$ WLOG, where $N = |G|$ and $N^\prime = |G^\prime|$ are the number of nodes in $G$ and $G'$ respectively. We apply local complementation (LC) on each node of $G$, for each such transformed graph, delete all possible groups of $N - N^\prime$ nodes, check if the resulting graph is LC-equivalent to $G^\prime$ (for which there is a polynomial time algorithm~\cite{van2004efficient}), and record if the $(G, G^\prime)$ pair are a vertex minor pair or not. The time complexity of this algorithm is:
 \begin{equation}\label{eq:4}
 O(N \times 2^{2N} \times 2^{-N^\prime}),
 \end{equation}
 For a derivation of the above, see Appendix \ref{appendix:bruteforce}.
 
\subsection{Constructing the feature vector}\label{sec:featurevector}

For our classical algorithm's feature vector, we calculate the eigenvalues of the Laplacians of each graph, concatenate and zero pad the arrays to even their size. For our quantum algorithm's feature vector, we concatenate the GBS samples for both graphs and zero pad the arrays to even their size. 
\subsection{Repeated trials}
  \begin{figure}
    \centering
    \includegraphics[width=0.8\linewidth]{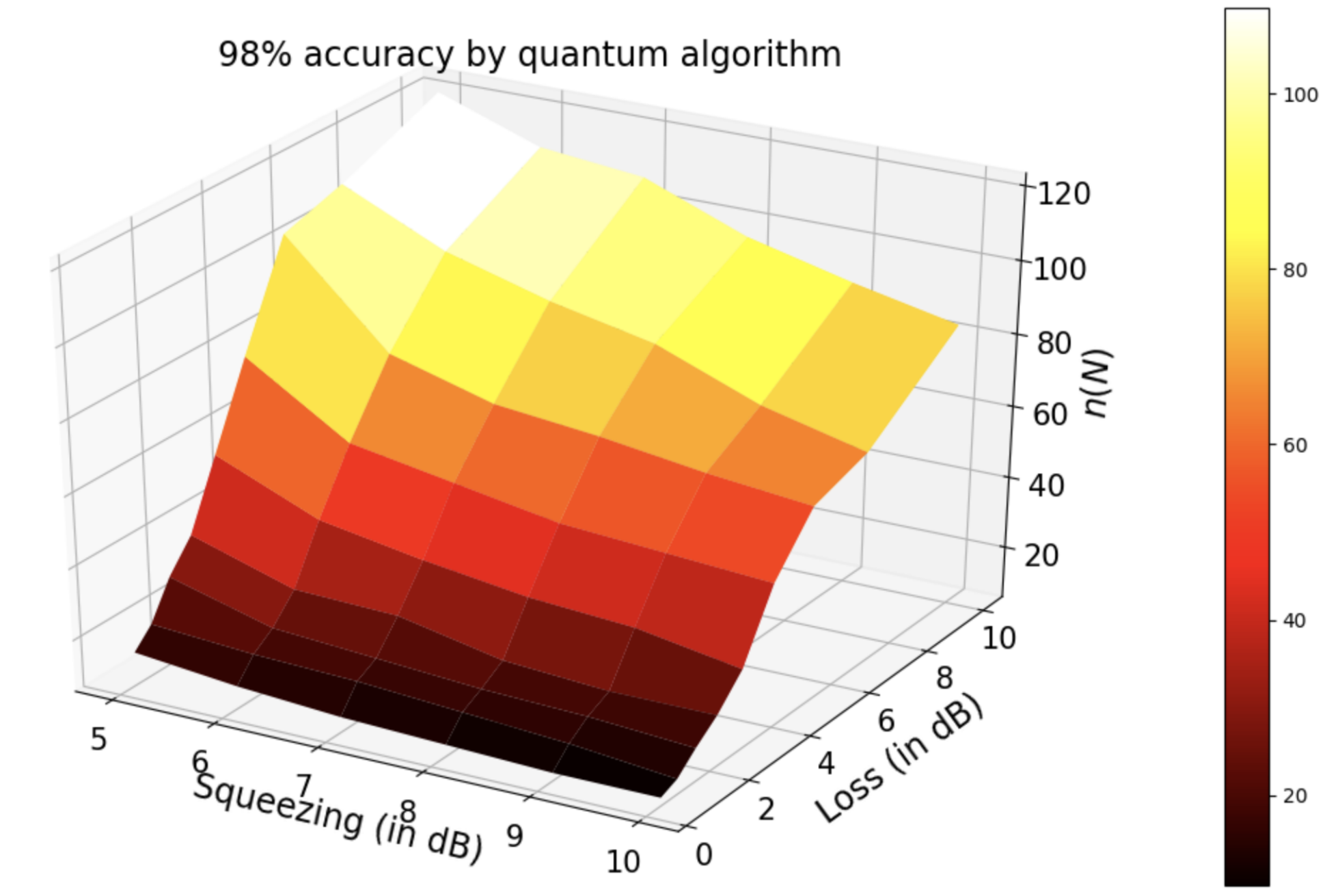}
    \caption{A 3D plot showing the value of $n(N)$ required to get a 98$\%$ accuracy by the quantum algorithm. For each value of squeezing and loss, we average over 7 graph pair's $n(N)$ value. We observe that as the value of maximum squeezing per mode increases, the value of $n(N)$ decreases and as loss increases, the value of $n(N)$ increases.}
    \label{fig:3D}
\end{figure}
In order to increase the accuracy of our machine learning algorithm to a desired level of 99\% we use repeated trials. It involves running the algorithm $n$ times (an odd number) and selecting the majority result as the final output, which is the maximum likelihood strategy assuming that each trial is statistically independent. If the probability of (classification) error in a single trial is $e$, then the probability of error after running the classifier $n$ times can be expressed as:
\begin{equation}
\label{eq:5}
  P_{\text{error}}(n,e)= \sum_{k=0}^{\lfloor \frac{n}{2} \rfloor}{n\choose k}e^{n-k}(1-e)^{k}.
\end{equation}
As shown in Appendix \ref{appendix:trials}, in the regime of low per-shot accuracy, i.e. $\epsilon \equiv 0.5 - e \ll 1$, the number of trials needed to achieve a given $P_{\text{error}}(n,e) \equiv \delta$, scales as:
\begin{equation}\label{eq:6}
    n \sim 1/\epsilon^2,
\end{equation}
where the constant in the above scaling is a function of $\delta$. The per-shot accuracy $\epsilon(N)$ is a function of the problem size $N$, i.e., the number of vertices in the graph(s), and is dependent on the algorithm (classical or quantum). The larger the problem size $N$, lower will in general be the per-shot accuracy $\epsilon(N)$, and larger will be the number of trials $n(N) \sim 1/\epsilon(N)^2$ needed to reach a desired overall error rate $\delta$. 

\begin{figure}[t]
  \centering

    \includegraphics[width=0.9\linewidth]{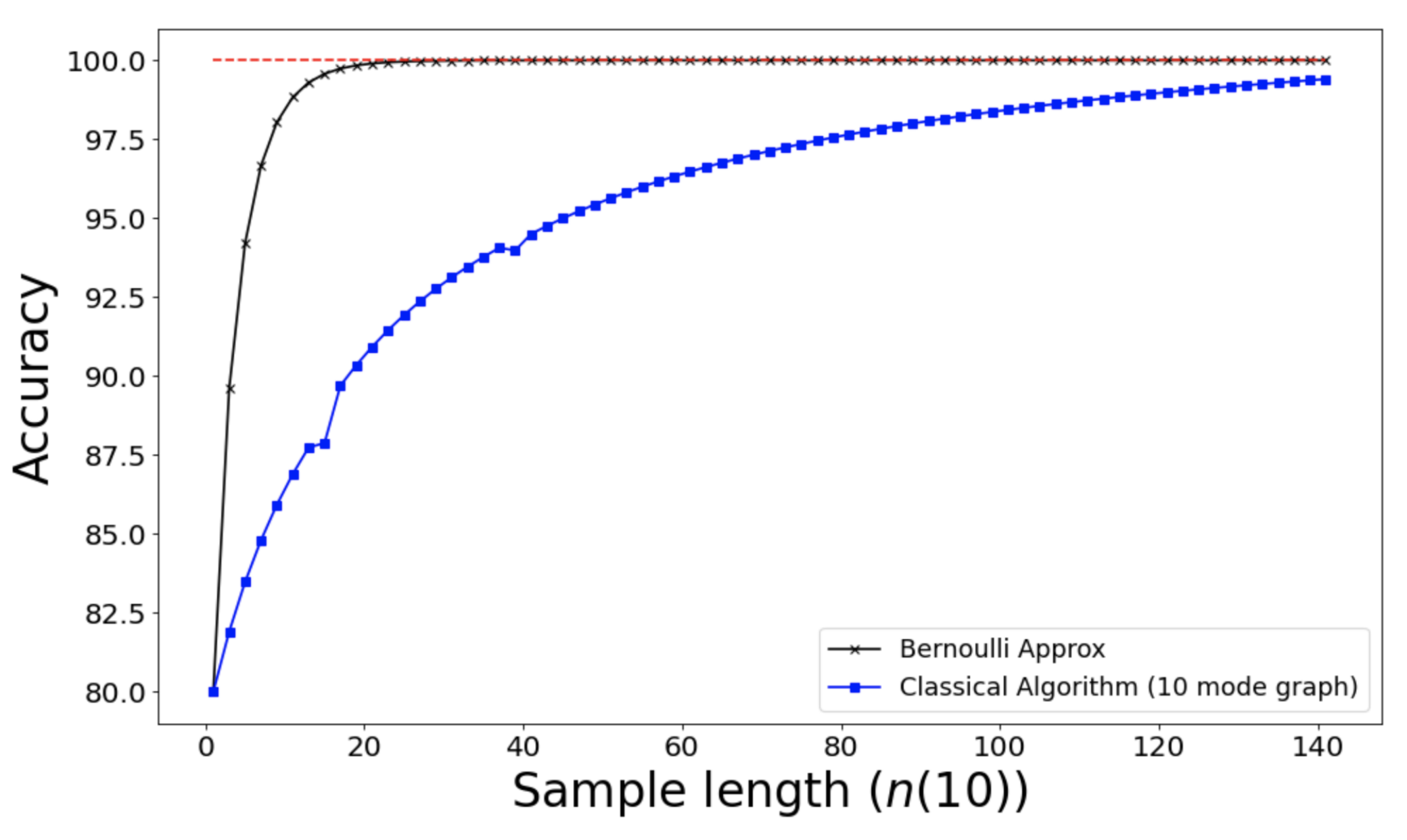}

    \caption{Accuracy of the classical algorithm vs $n(N)$, where $N$ = 10. The accuracy is computed by averaging over 7 graph pairs for each particular $n(N)$.
    In an ideal scenario where each trial is independent of the others, the accuracy would follow the Bernoulli approximation, resulting in increased accuracy with higher values of $n(N)$. However, due to the negative correlations present in the algorithm, it does not perform optimally and approaches a near-perfect accuracy but falls short of reaching 100$\%$.}
    \label{fig:classical}
\end{figure}

Note that the above argument relies on the $n$ trials to be i.i.d., which in general is not true, resulting in deviations from the above scaling, as seen in Fig.\ref{fig:classical} and Fig.\ref{fig:quantum}. 

\subsection{Propagation in programmable photonic integrated circuit}
Embedding a graph of $N$ nodes in a GBS requires an $N$-mode interferometer. The design of a general $N$-mode passive linear-optical unitary as a multiport interferometer~\cite{clements2016optimal}, needs depth $N$. In other words, the time it takes an optical pulse to traverse from the input to the output $t_{\rm o} =O(N)$, as it must go through $N$ two-port beamsplitters. With reasonable assumptions on modern integrated photonic devices,
 \begin{equation}\label{eq:7}
t_{\rm o} \approx {1.44 \times N \times 10^{-13}}/{3} \,\,{\text{seconds}},
 \end{equation}
where we have used reasonable assumptions of each beam splitter having a length of 10 $\mu$m, speed of light in vacuum $\approx 3 \times 10^8$ m/s, and refractive index of silicon waveguide $1.44$.

Another factor at play is the repetition period (in seconds) of squeezed-light pulses sent into a GBS,
  \begin{equation}\label{eq:8}
t_{\rm r}= 1/R,
 \end{equation}
where $R \sim$ MHz is the pulse-repetition rate of the pump laser. 

 \subsection{Loss}
  \begin{figure}[t]
  \centering

    \includegraphics[width=0.9\linewidth]{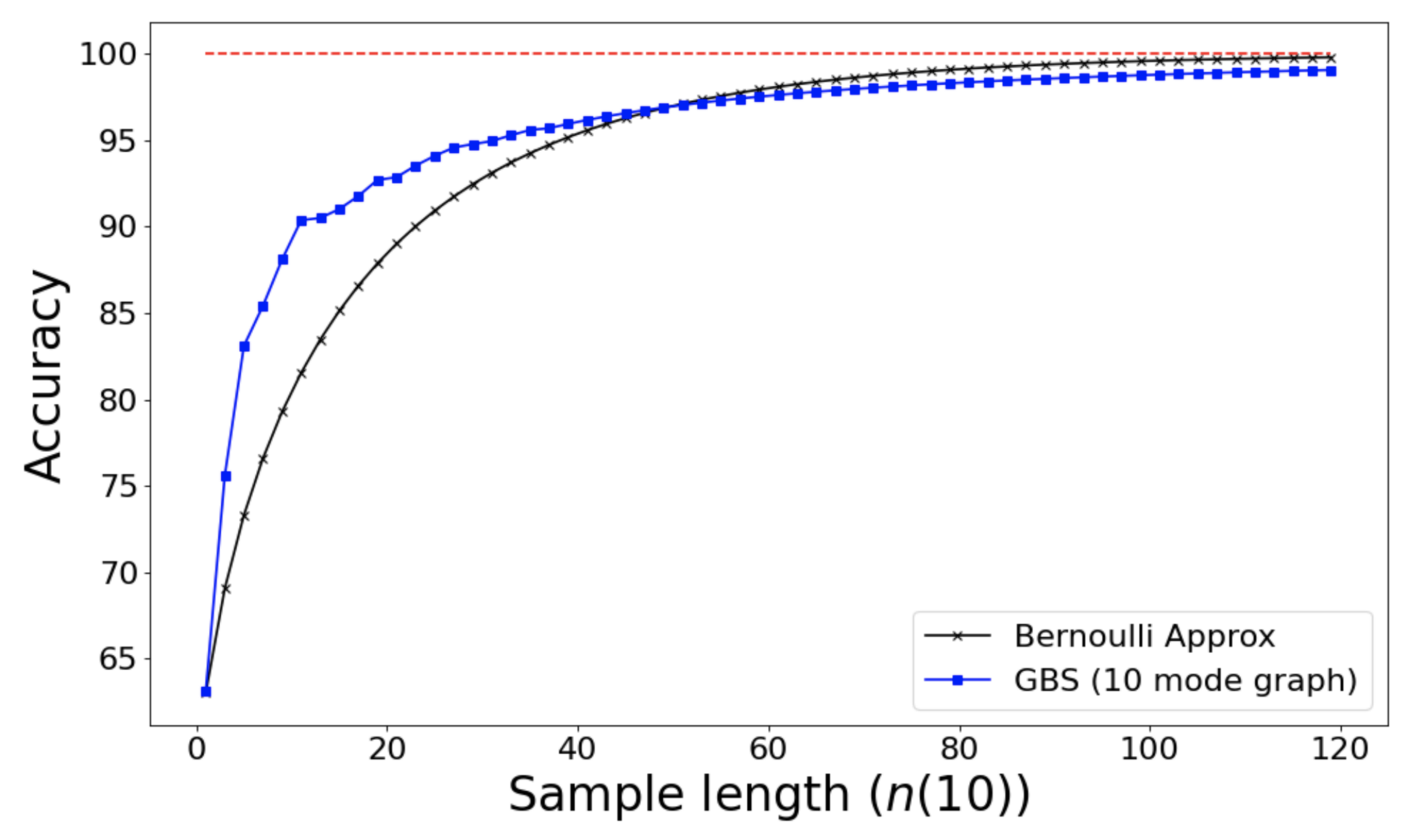}

    \caption{Accuracy of the quantum algorithm vs $n(N)$, where $N$ = 10. The accuracy is computed by averaging over 7 graph pairs for each particular $n(N)$. In an ideal scenario where each trial is independent of the others, the accuracy would follow the Bernoulli approximation, resulting in increased accuracy with higher values of $n(N)$. However, we observe that there are some positive correlations until 47 and then the Bernoulli takes over due to the negative correlations present in the algorithm and approaches a near-perfect accuracy but falls short of reaching 100$\%$.}
    \label{fig:quantum}
\end{figure}
 Three main sources of loss in a GBS are: (1) optical coupling efficiency (into and out of the chip), (2) on-chip losses, and (3) detector efficiency. Let us denote the effective transmissivity contributed by the overall coupling efficiency by $\eta_{\rm c}$, that of on-chip losses by $\eta_{\rm o}$, and that of the detector efficiency by $\eta_{\rm d}$. If the GBS is designed to output a state with covariance matrix $\sigma_{\rm pure}$ when there are no losses (Eq.~\eqref{eq:2}), the covariance matrix of the state produced by the lossy GBS,
 \begin{equation}\label{eq:9}
     \sigma_{\text{lossy}}= \eta_{\rm c}\eta_{\rm o}^N\eta_{\rm d}\sigma_{\text{pure}} + (1-\eta_{\rm c}\eta_{\rm o}^N\eta_{\rm d})I.
 \end{equation}
 Please see Appendix \ref{appendix:loss} for the derivation.

\section{Algorithms}
 
 \subsection{Randomized Classical Algorithm}

We first propose a randomized classical algorithm that uses the graph spectra, LCs, and $n$ repeated trials. Given graphs $G_1$ and $G_2$, where $|G_2| \le |G_1|$, we construct a list of graphs $G_3^{(m)}$, $1 \le m \le n$, by performing LC on random subsets of vertices of graph $G_1$, thereby ensuring $G_3^{(m)}$ is LC equivalent to $G_1$, $\forall m$. If $G_2$ is a vertex-minor of $G_1$, $G_2$ will also be a vertex minor of $G_3$ and vice versa (see Appendix \ref{appendix:random}). For each trial $m$, we use the spectral method to test $G_3^{(m)}$ and $G_2$ for the vertex minor test, by constructing the classical feature vector using the method outlined in Section~\ref{sec:featurevector}. We use a majority vote over $n$ trials, to declare the final verdict. The number of trials needed depends upon the desired accuracy. Clearly, the $n$ trials are not i.i.d. due to correlations between trials, and hence the algorithm's overall accuracy starts to plateau at a fixed value, as shown in Fig.~\ref{fig:classical}. Our classical algorithm first finds a locally equivalent graph of the larger graph, constructs its Laplacian Matrix and then finds its corresponding eigenvalues followed by the method of repeated trials to increase our accuracy. The time complexity of deriving a random local equivalent graph is $O(N^2)$. The time complexity to construct a Laplacian of an $N$ node graph is $O(N^2+M)$ where $M = O(N^2)$ is the number of edges in the graph. The time complexity of calculating the eigenvalues of a $N \times N$ matrix is $O(N^3)$ using the QR decomposition, and that of predicting using a linear SVM is $O(N)$. Given that the number of repeated trials needed for 97$\%$ accuracy ($\delta = 0.03$) is $n_{\rm c}(N)$ (which is smaller than would be needed by the i.i.d. trials assumption with the same one-shot accuracy as our algorithm, as shown in Fig.~\ref{fig:classical}), the time-complexity of our classical algorithm is given by:
 \begin{eqnarray}
     \label{eq:10}
 T_{\rm c}&=&n_{\rm c}(N) \times (O(N^3)+O(N^2)+O(N)) \nonumber \\
 &=&n_{\rm c}(N) \times O(N^3).
 \end{eqnarray}


 \subsection{Quantum Algorithm}

Given graphs $G_1$ and $G_2$, where $|G_2| \le |G_1|$, we calculate the quantum feature vector (as described in Section~\ref{sec:featurevector}) and give it as an input to our SVM algorithm. We take $n(N)$ samples until we reach a desired accuracy (say, 99$\%$, i.e., $\delta = 0.01$). As we see in Fig.\ref{fig:quantum}, because of correlations from shot to shot (since it is the same graphs encoded in the GBS), the large-$n(N)$ accuracy is lower than that predicted by the i.i.d. Bernoulli model, as in the classical case. For the quantum realization, we embed the graph in our GBS using the Takagi decomposition of the graph's adjacency matrix, to give us the squeezing parameters and the $N$-mode unitary of the GBS. The time complexity for the Takagi decomposition is $O(N^{2})$ \cite{xu2009twisted}, which is better than all other methods.
Given that the number of repeated trials needed to get a 97$\%$ accuracy (i.e., $\delta = 0.03$) is $n_{\rm q}(N)$ (which is smaller than that would be needed by the hypothetical case of i.i.d. trials with the same one-shot accuracy as our algorithm, as shown in Fig.~\ref{fig:quantum}), the full time-complexity of the classical realization of this algorithm, i.e., when simulated on a classical computer, is:
\begin{equation}
 T_{\text{cgbs}}=n_{\rm q}(N) \times
  O(N^42^{N/2}),
    \label{eq:11}
\end{equation}
and that of its quantum realization (using a GBS device) is:
  \begin{align}
  \label{eq:12}
  \begin{split}
  T_\text{qgbs}=O(N^2) + n_{\rm q}(N).
  \end{split}
\end{align}

 \section{Results} 
Here we show that: (1) our classical algorithm based on randomized LCs and graph spectra performs better compared with other well-known graph classification methods; and (2) our GBS-based quantum algorithm performs better than the classical method with an integrated-photonic realization for reasonable assumptions on loss and squeezing.

\subsection{Comparing classical algorithm's one-shot accuracy}
\begin{figure}[t]
  \centering

    \includegraphics[width=0.9\linewidth]{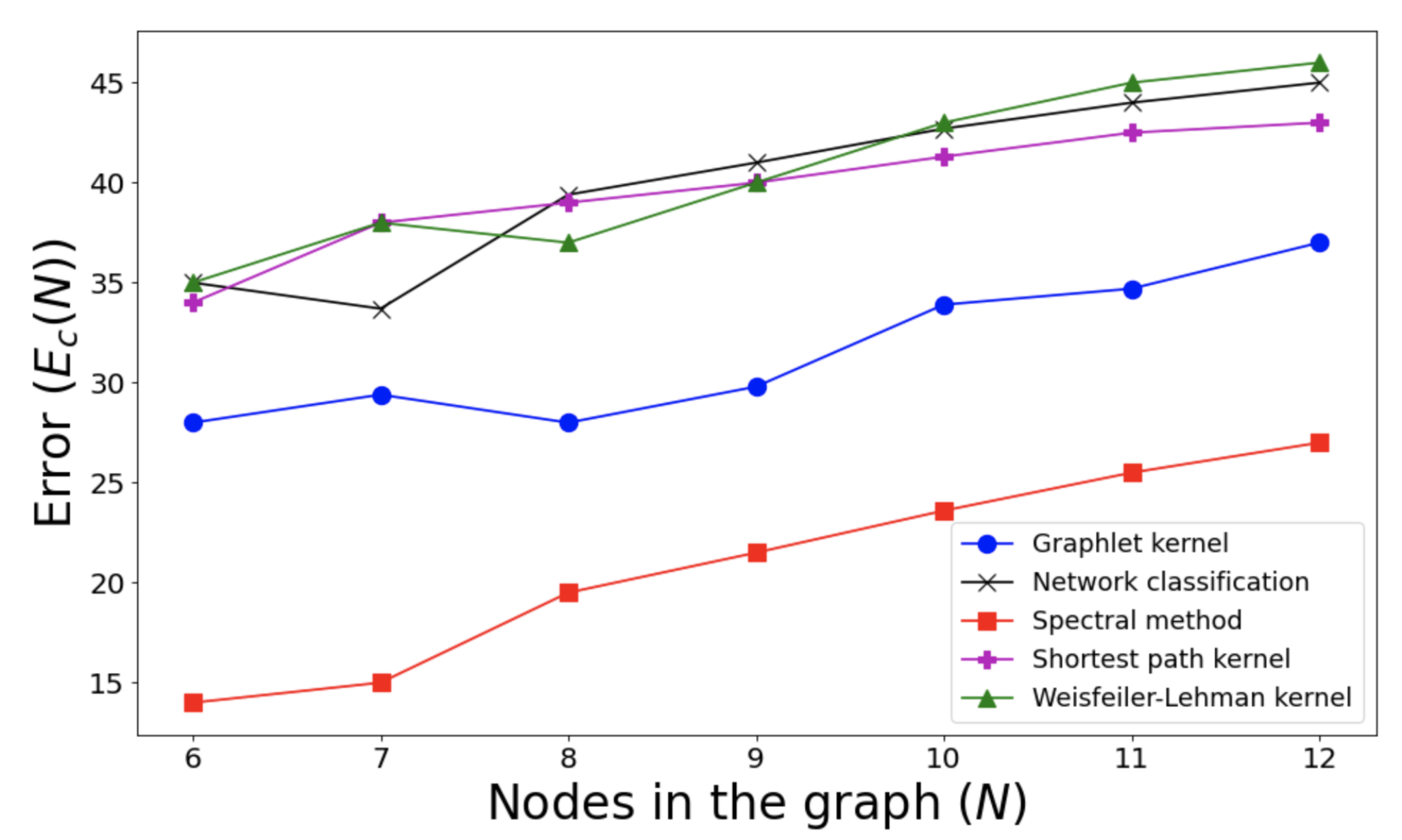}

    \caption{Error comparison of the classical algorithms, averaged over 1000 datasets. We see that the spectral method performs better than the other well-known graph similarity algorithms.} 
    \label{fig:error}
\end{figure}

We first create our dataset of graphs. For each $N$, number of nodes in the graph, we generate $500$ ``vertex-minor" graph pairs and $500$ ``not vertex-minor" graph pairs, using the brute force method. For the Graphlet, Shortest path and WL Kernel methods, we use a C-supported Support Vector Machine (SVM) by passing through the precomputed kernels. We use 25$\%$ of our dataset as test data and the remaining for training our SVM. The C-parameter of the SVM controls the penalty for misclassifying training examples and we use a C value in the range 5 to 10.  The best model is then used to get the accuracy of the test set. 
The hyperparameters used for Network Classification are written in Appendix \ref{appendix:classical}. For the spectral method, we use a linear SVM with the same parameters as C-supported SVM. The best model is then used to get the accuracy of the test set. After plotting the error versus the nodes in the graph (Fig.\ref{fig:error}), for each classical algorithm, we see that our spectral method is indeed the best classical algorithm among the ones we evaluated. 
 
\subsection{Comparing our classical and quantum algorithms}
\begin{figure}[t]
  \centering
    \includegraphics[width=0.9\linewidth]{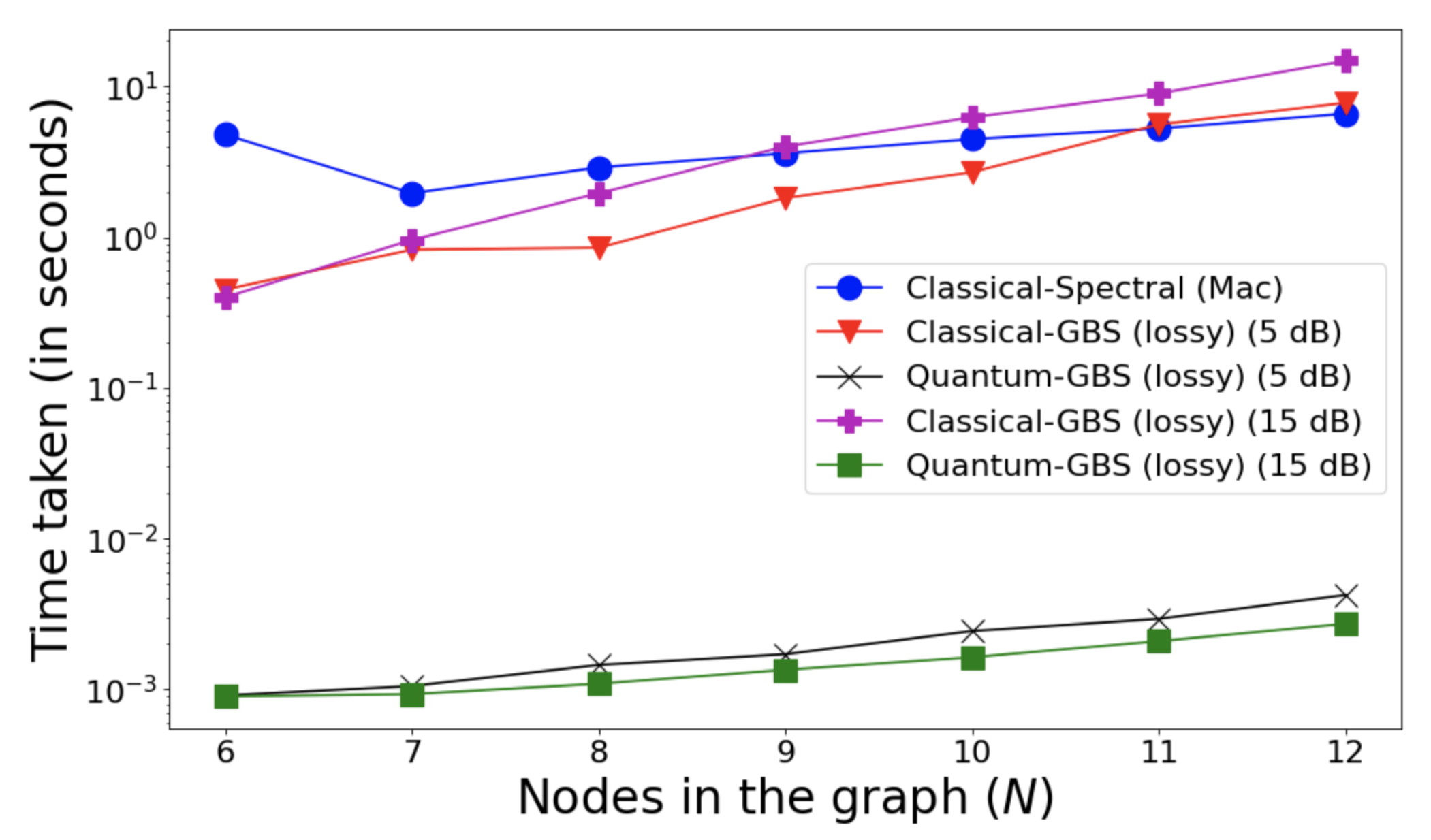}
    \caption{For all quantum GBS simulations, the photon loss assumed is $\approx 1.2$ dB. The quantum GBS realization with $5$ dB squeezing per mode performs $10^3$ times better than the classical algorithm (run on the most powerful MacBook Pro). Increasing squeezing improved performance.} 
    \label{fig:compare}
\end{figure}

Finally, we demonstrate that with fair assumptions on losses, amount of squeezing per mode and squeezed-pulse repetition rate, our quantum algorithm performs up to $10^3$ times better than the best classical algorithm discussed above, averaged over the same dataset of graph pairs (Fig. \ref{fig:compare}), with the classical algorithm running on a powerful classical computer. For our GBS simulation, we assumed $80$$\%$ coupling efficiency (in and out), $95$$\%$ detector efficiency and $0.25$ dB/cm of on-chip losses. Maximum squeezing per mode is taken to be $5$ dB and a squeezed-pulse repetition rate of $10$ MHz. 

A likely explanation might be that the Takagi decomposition, which drives the dominating term of the GBS algorithm's complexity, scales faster than calculating eigenvalues of the Laplacian of the graphs, hence performing better even though the GBS requires more trials than a classical algorithm. 

\section{Conclusion}
We propose a hybrid quantum-classical approach using a Gaussian Boson Sampler (GBS) to solving the NP-Complete decision problem of determining if two graphs are a vertex-minor pair (or not). The premise behind our approach is the claim that by mapping a graph's adjacency matrix into the quantum state of a GBS, the classically-hard-to-sample photon-click-pattern samples produced by the GBS `powerfully' encode features of the graph embedded in it, which in turns assists with this classification problem. We design a graph embedding, extending work of~\cite{schuld2019quantum}, that lets us pick a tunable squeezing amount at the GBS input (a hard-to-produce quantum optical resource) at the expense of a lower one-shot classification accuracy, which in turn results in a larger number of repeated trails to obtain a desired (high) target accuracy, and show that repeated trials do get us to very high accuracies despite the inherent shot-to-shot correlations. We also propose a classical spectral algorithm for the vertex minor problem that uses randomized generation of an LC-equivalent graph library of one of the two graphs presented, followed by a support vector machine (SVM) classifier fed with a feature vector formed using a zero-padded set of the graph Laplacians' eigenvalues, which we show outperforms various state-of-the-art classical graph classification algorithms. A rigorous time-complexity analysis of our algorithms is hard given the underlying scaling of the one-shot error rate with the problem size is difficult to derive. However, using numerical analyses, we show that with a readily-realizable on-chip GBS, our hybrid special-purpose quantum processor would outperform the aforesaid classical algorithm when executed on the most powerful MacBook Pro currently available. 

Many questions remain open, including the possibly better embedding of this problem to a GBS (or another NISQ quantum processors), finding possibly better classical algorithms than the one we presented, or using a different machine learning (ML) classifier back-end that might give higher accuracy. Future work also entails studies of other NISQ processors for various practical problems in optimization, search, estimation and classification (e.g., of chemical vibronic spectra), and other graph problems, and performance comparisons under realistic assumptions with the respective best-known classical algorithmic approaches. Finally, progress in rigorous analyses of time complexity scaling of random-sampling based NISQ algorithms will put this field on a much stronger footing. 


\section*{Acknowledgment}

The authors would like to thank Ashlesha Patil for suggesting the Vertex Minor problem to study in this NISQ context, and Don Towsley, Nathan Killoran and Prithwish Basu for insightful discussions and suggestions. This research was supported by the UArizona FoTO scholarship, and a DoE project awarded under grant number 4000178321.

\bibliographystyle{ieeetr}

\bibliography{citations} 
\newpage

\begin{appendices}

\section{Derivation for repeated trials}\label{appendix:trials}

Here we focus on deriving how our number of trials $n$ is related to the accuracy of our ML algorithms. The probability of error after running the classifier $n$ times can be expressed as a function of $e$ and $n$ as a partial sum of the binomial distribution,
\begin{equation}\label{eq:13}
  P_\text{{\text{error}}}(n,e)= \sum_{k=0}^{\lfloor \frac{n}{2} \rfloor}{n\choose k}e^{n-k}(1-e)^{k},
\end{equation}
which approximates to the normal distribution when $n$ is large:
\begin{equation}\label{eq:14}
   {n\choose k}e^{n-k}(1-e)^{k} \approx \frac{1}{\sigma\sqrt{2\pi}}{e^{-(k-\mu)^{2}/(2\sigma^{2})}},
\end{equation}
where $\mu=n(1-e)$ and $\sigma^2=ne(1-e)$. Next, we will derive an expression for $n$ for a desired overall $P_{\text{error}} \equiv \delta$.

\begin{equation}\label{eq:15}
    \delta(n,e)= \sum_{k=0}^{\lfloor \frac{n}{2} \rfloor}{n\choose k}e^{n-k}(1-e)^{k}, \, {\text{hence}},
\end{equation}
\begin{equation}\label{eq:16}
   \delta(n,e) \approx \frac{1}{\sigma\sqrt{2\pi}}\int_{-\infty}^{\lfloor \frac{n}{2} \rfloor} {e^{-(x-\mu)^{2}/(2\sigma^{2})}} \,dx.
\end{equation}
Substituting $y=x/n$, we get
\begin{eqnarray}\label{eq:17}
    \delta(n,e)&\approx& \frac{n}{\sigma\sqrt{2\pi}}\int_{-\infty}^{\frac{1}{n}\lfloor \frac{n}{2} \rfloor} {e^{-(ny-\mu)^{2}/(2\sigma^{2})}} \,dy,\\
    &=& \frac{\sqrt{n}}{\sqrt{2e(1-e)}}\int_{-\infty}^{\frac{1}{n}\lfloor \frac{n}{2} \rfloor} {e^{\frac{-n(y-(1-e))^{2}}{2e(1-e)}}} \,dy\label{eq:18} \\
&=& \frac{1}{\sigma\sqrt{2\pi}}\int_{-\infty}^{\mu} {e^{-(x-\mu)^{2}/(2\sigma^{2})}} \,dx \nonumber \\
    &-& \frac{1}{\sigma\sqrt{2\pi}}\int_{\lfloor \frac{n}{2} \rfloor}^{\mu} {e^{-(x-\mu)^{2}/(2\sigma^{2})}} \,dx.\label{eq:19}
    \end{eqnarray}
The first integral in the last line above equals $\frac{1}{2}$. Substituting $y=x-\mu$ in the second integral and swapping the limits, we get:
\begin{eqnarray}\label{eq:20}
    \delta(n,e)&\approx& \frac{1}{2} + \frac{1}{\sigma\sqrt{2\pi}}\int_{0}^{(\lfloor \frac{n}{2} \rfloor- \mu)} {e^{-(y)^{2}/(2\sigma^{2})}} \,dy, \\
    &=&\frac{1-{\rm erf}(k)}{2},\label{eq:21}
\end{eqnarray}
where ${\rm erf}(z)= \frac{2}{\sqrt{\pi}}\int_{0}^{z}{e^{-t^2}} \,dt$, with
\begin{equation}
\label{eq:22}
    k(n,e)=\frac{(1-e)n-\lfloor \frac{n}{2} \rfloor}{\sqrt{2ne(1-e)}}.
\end{equation}
We know that erfc$(z)$=$1$-erf$(z)$, hence,
\begin{equation}\label{eq:23}
    2\delta = {\rm{erfc}}(k) \approx \frac{e^{-k^2}}{\sqrt{\pi}k}, \,{\text{for}}\, k \gg 1.
\end{equation}
Taking natural logarithms on both sides,
\begin{equation}
\label{eq:24}
    -k^2 - \ln k -0.5724= \ln 2\delta.
\end{equation}
Let us express the solution of the transcendental equation above, as $k(\delta)$.
Using Eq.\ref{eq:22} we get,
\begin{equation}
\label{eq:25}
    \frac{(1-e)n-\lfloor \frac{n}{2} \rfloor}{\sqrt{2ne(1-e)}}=k(\delta).
\end{equation}

Assuming $\lfloor \frac{n}{2} \rfloor=\frac{n}{2}$ for $n\gg1$ and squaring both sides, we get

\begin{equation}
\label{eq:26}
    \frac{n(0.5-e)^2}{2e(1-e)}=k^2(\delta).
\end{equation}

and assuming that the one-shot accuracy $\epsilon= (0.5-e) \ll 1$, we get 
\begin{equation}\label{eq:27}
    n \approx k(\delta)^2/2\epsilon^2,
\end{equation}
establishing the scaling 
\begin{equation}\label{eq:28}
    n \sim \frac{1}{\epsilon^2},
\end{equation}
assuming i.i.d. trials. Note that the i.i.d. trials assumption is not true for any of the algorithms studied in this paper, hence this scaling provides a bound.

Using the Newton-Raphson method and substituting a specific value of the overall target accuracy after $n$ trials, $\delta=0.01$ in Eq.\ref{eq:24}, we get $k= 1.6796$, substituting which in Eq.~\eqref{eq:27}, we get:
\begin{equation}\label{eq:29}
    n(e) \approx 1.41/\epsilon^2.
\end{equation}

The value of $\epsilon$ depends on $N$, the number of vertices in a graph, and the choice of the algorithm. Hence the required number of trials $n$ also depends upon $N$.

\section{Classical Algorithms used}\label{appendix:classical}
We are assuming $G$ and $G'$ as our two graphs of comparison where the number of nodes of $G'$ are less than or equal to $G$. 
\subsection{Graphlet Sampling Kernel}
  \begin{figure}[t]
    \centering
    \includegraphics[width=0.99\linewidth]{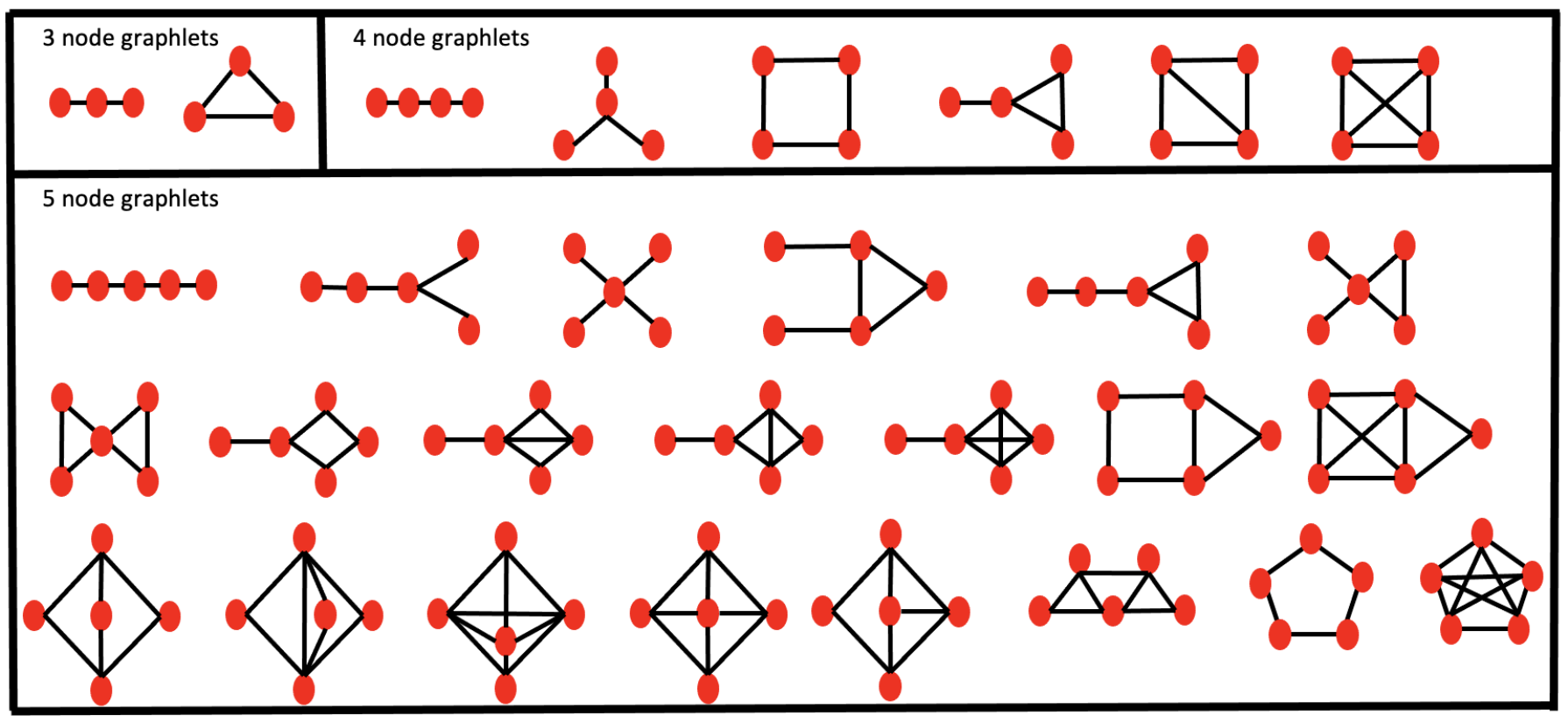}
    \caption{All $k \in \{3,4,5\}$ connected non-isomorphic graphlets we used for our Graphlet Sampling Kernel algorithm.}
    \label{fig:graphlets}
\end{figure}
The Graphlet Sampling Kernel \cite{shervashidze2009efficient} is a popular machine-learning-based method for graph classification and clustering tasks. The calculation of feature vectors proceeds as follows.
\\
Let $S_k \equiv \{H_1, \ldots, H_{r_k}\}$ denote the set of all ($r_k$) distinct size-$k$ graphlets. For example, Fig.~\ref{fig:graphlets} sketches the elements of $S_3$, $S_4$ and $S_5$. Let us denote the two input graphs to our vertex minor problem as $G_1$ and $G_2$, with $N_1 > N_2$, with $N_i = |G_i|$, $i = 1, 2$. For graph $G_i$, let us define a set $T_{i,k,m}$ to contain all instances of graphlet $H_m$ of size $k$ in graph $G_i$, where $i \in \{1,2\}$, $m \in \{1, \ldots, r_k\}$, and $k \in \{3, 4, \ldots, N_2-1\}$. Let us define: 
\begin{equation}\label{eq:30}
    f_{i,k,m}= |T_{i,k,m}|,
\end{equation}
and the concatenated vectors:
\begin{equation}\label{eq:31}
    f_{i,k} \equiv \left[f_{i,k,1}, \ldots, f_{i,k,r_k}\right],
\end{equation}
and
\begin{equation}\label{eq:32}
    f_{i} \equiv \left[f_{i,3}, \ldots, f_{i,N_2-1}\right].
\end{equation}
Finally, we define the feature vector for graph $G_i$, 
\begin{equation}\label{eq:33}
    L_i = \frac{f_i}{\sum_{k=3}^{N_2-1}\sum_{m=1}^{r_k}f_{i,k,m}}.
\end{equation}

We then calculate the kernel given by the formula:
\begin{equation}\label{eq:34}
    k_\text{GS}(G_1,G_2)= L^{T}_1L_2.
\end{equation}
For training, we then use this precomputed kernel and pass it through our SVM classifier labelling it ``vertex-minor" or ``not vertex-minor" depending on the pair of graphs chosen. 

For our problem, we choose all possible non-isomorphic graphlets of size $k \in \{3,4,5\}$ giving us a total of 29 graphlets (shown in Fig.\ref{fig:graphlets}) because our $N_2$ has a minimum value of 6. 
\begin{figure}[t]
    \centering
    \includegraphics[width=0.99\linewidth]{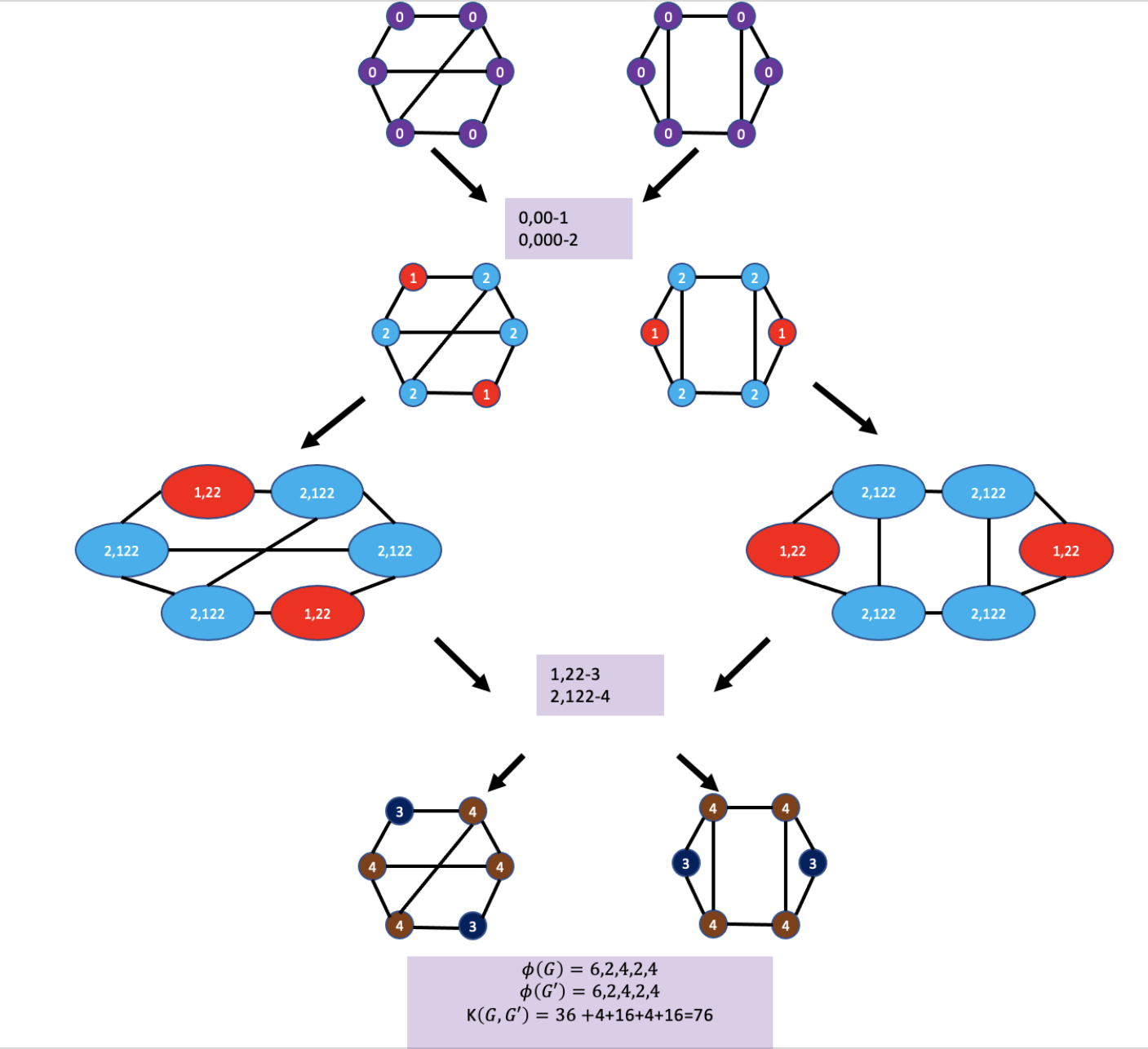}
    \caption{Example of the Weisfeiler-Lehman algorithm. The labels are updated by concatenating the vertex number with the adjacent vertices. The label is again updated by a pre-defined hash table. In the end, the vertex (defined in the hash table) frequencies become a feature vector and the kernel is defined as an inner product of the two vectors.}
    \label{fig:WL}
\end{figure}
\subsection{Shortest path Kernel}
The shortest-path kernel \cite{borgwardt2005shortest} is a graph analysis technique that involves breaking down complex graphs into their constituent shortest paths and then comparing these paths based on their lengths and the labels of their start and end points. Firstly, we would want to compute the all-pairs-shortest path for $G$ and $G'$ using the Floyd-Warshall algorithm. The Floyd-Warshall algorithm is a dynamic programming algorithm for finding the shortest path between all pairs of vertices in a graph. It achieves this by computing a matrix of shortest path distances, where each entry in the matrix represents the shortest path distance between two vertices. The shortest-path kernel is then given by:-
\begin{equation}\label{eq:35}
    k_\text{SP}(G,G')= \sum_{\substack{v_i, v_j \in G}}\sum_{\substack{v'_k, v'_l \in G'}} k(d(v_i,v_j), d(v_{k'},v_{l'})),
\end{equation}
where $d(u,v)$ gives us the shortest path between nodes u and v and k is the base kernel. For our problem, we used the linear kernel $k(d(v_i,v_j), d(v_{k'},v_{l'}))=d(v_i,v_j)*d(v_{k'},v_{l'})$. This is a precomputed kernel so we pass it through our SVM classifier labelling it ”vertex-minor” or ”not vertex-minor” depending on the pair of graphs chosen.

\subsection{Weisfeiler-Lehman Kernel}

  \begin{figure}[t]
    \centering
    \includegraphics[width=0.99\linewidth]{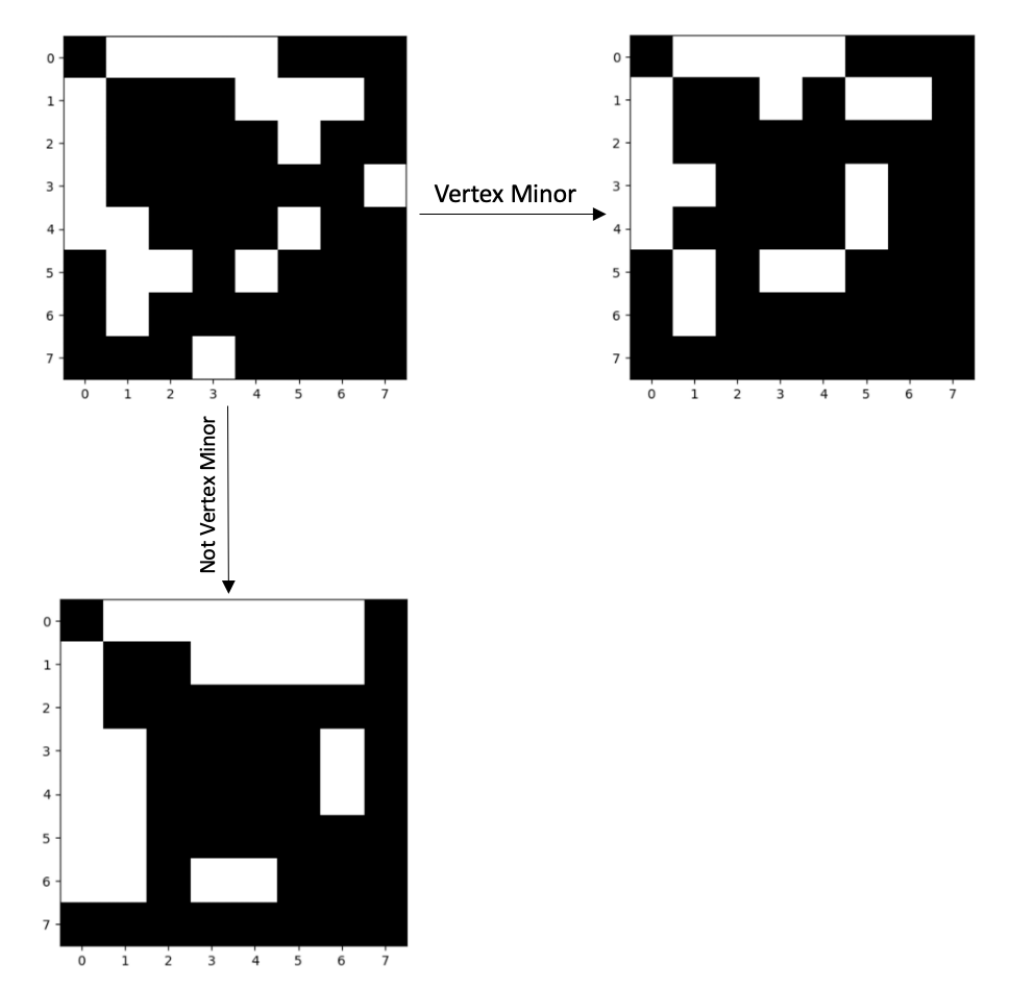}
    \caption{Network Classification Using Adjacency Matrix Embeddings for our vertex minor graphs. As we can see from the image transformation of the graphs, the vertex-minor pair looks similar compared to the not vertex-minor pair hence helping the neural network to predict the correct label.}
    \label{fig:network}
\end{figure}
The Weisfeiler-Lehman algorithm \cite{shervashidze2011weisfeiler} is a technique used to label each vertex of a graph with a multiset label. This label consists of the original vertex label and the sorted set of labels of its neighbouring vertices. The multiset label is then compressed into a new, shorter label. This relabeling process is repeated for several iterations. It is important to note that this procedure is applied simultaneously to all input graphs. If two vertices from different graphs have identical multiset labels, they will receive the same new label.
\\
\begin{figure*}
    \centering
    \subfigure[ ]
    {
        \includegraphics[width=1.0\linewidth]{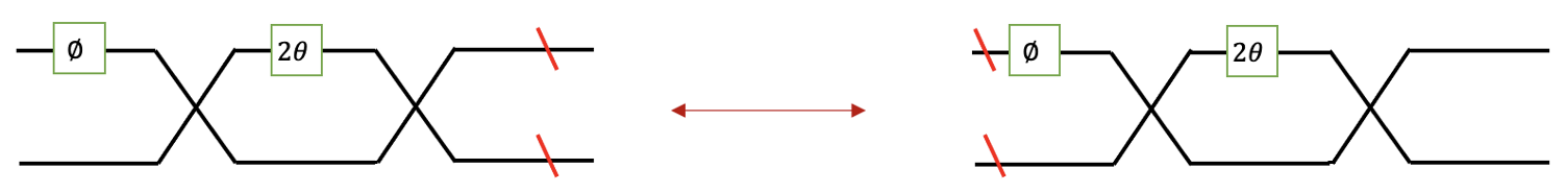}
        \label{fig:first_sub}
    }
    \\
    \subfigure[ ]
    {
        \includegraphics[width=1.0\linewidth]{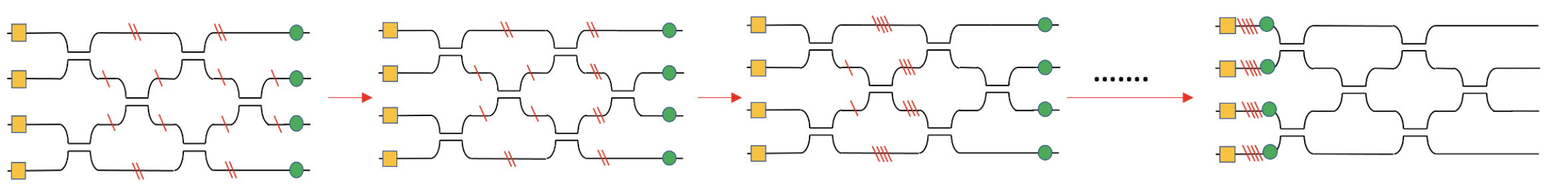}
        \label{fig:second_sub}
    }
    \caption{Coupling the loss terms together in the input mode of the multimode interferometer. (a) If there is the same loss in both the output mode on an MZ interferometer, it is equivalent to having the same loss in both the input modes too. (b) Here we show how the loss terms move back in the interferometer. A double dash signifies twice the loss because the length of the fiber there is double that of the single dash.}
    \label{fig:loss}
\end{figure*}


Formally, we assign the same label, $l=l_0$ to all the vertices of our graphs $G$ and $G'$. In the next iteration, each label is updated with a sequence of its own label followed by labels from the neighbouring vertices. We rename this sequence of labels for each vertex with a new label $l_1=h(l_0)$,  from a defined hash table, $h$ and repeat this process for a certain number of iterations. The relabeled graphs can be called as $G_{i}$ where $i$ is the number of iterations $G$ and then we have an array of graphs $\{G_0,G_1,G_2....G_h\}$ where h is the total number of iterations. We then define our kernel as:-
\begin{equation}\label{eq:36}
    k_{WL}(G,G')= k(G_0,G'_0)+k(G_1,G'_1)+...+k(G_h,G'_h)
\end{equation}
where $k$ is our base kernel. We usually count the number of nodes corresponding to each label defined in the hash table, make a vector for each graph and then take an inner product of the two vectors. This is a precomputed kernel so we pass it through our SVM classifier labelling it ``vertex-minor" or ``not vertex-minor" depending on the pair of graphs chosen. An example of this kernel calculation is given in Fig: \ref{fig:WL}.

\subsection{Network Classification Using Adjacency Matrix Embeddings}
Network classification using adjacency matrix embeddings \cite{wu2016network} and deep learning is an algorithm where the adjacency matrix of the graph is embedded into an image and we perform image classification using a neural network. The procedure for embedding the adjacency matrix of a graph into an image for the neural network is given below:-

\begin{enumerate}

    \item \emph{Ordering of Nodes}- To make the adjacency matrix informative, an ordering algorithm is used. This algorithm starts by selecting the node with the highest degree, and if there is a tie, it chooses the node with the largest k-neighbourhood for increasing values of k. The algorithm then proceeds to order the rest of the nodes by their shortest path distances to the already ordered nodes. If a tie exists, it prioritizes the node with the highest degree or picks randomly. This ordering algorithm prioritizes nodes that are well-connected and closer to the previously ordered nodes, while also considering the degree and size of each node's neighbourhood.
    \item \emph{Rescale Algorithm}- Since we are working with graphs of different sizes, we need to rescale all the graphs to one particular dimension of the image. There are two ways of doing that, image resizing and padding. In image resizing, the ``pixels" of the matrix are mapped linearly to the corresponding pixels in the final resized image. This mapping ensures that the information in the adjacency matrix is preserved even after resizing, allowing for accurate analysis and comparison of graphs of different sizes. In padding,  zeros are added as padding to the matrix, with the original matrix remaining in the top-left corner. This padding ensures that the matrix is resized uniformly without changing the relative positions of the nodes or altering the overall connectivity of the graph. For our algorithm, we use padding since resizing requires us to resize each graph to a new graph whose dimension is the least common multiple of both the dimensions of the graph (which is a large number) and hence cannot be encoded efficiently.
\end{enumerate}
After ordering the nodes and padding to a particular size, we flatten the pixel matrices and concatenate the flattened arrays of the pair of graphs depending if they are vertex minor or not. We then pass these arrays into our neural network with three layers. The first layer has 64 neurons and uses the rectified linear unit (ReLU) activation function. The second layer has 240 neurons and also uses the ReLU activation function. The third layer has 20 neurons and uses the rectified linear unit (ReLU) activation function. The final layer has a number of neurons equal to the number of classes in the dataset and uses the softmax activation function, which normalizes the output of the network to represent the probabilities of each class. The autoencoder uses a learning rate of 0.001 during pre-training and a learning rate of 0.1 during fine-tuning. To determine when to stop the training process, 10\% of the training data is used as a validation set. Pre-training uses a maximum of 10 epochs, while fine-tuning uses a maximum of 400 epochs. Both pre-training and fine-tuning use cross-entropy cost. As we can see from Fig.\ref{fig:network}, the vertex-minor pair of graphs look similar to the not-vertex minor pair. 

\section{Brute force algorithm to train SVM backend}\label{appendix:bruteforce}

To train the SVM back end, both for our classical (spectral) method as well as the GBS method, we generated a data set comprising $500$ vertex-minor graph pairs, and $500$ non-vertex-minor graph pairs, by brute force. For this, we need an exact algorithm (zero missed detections or false alarms) to determine if two graphs are related to each other by a series of local complementations and vertex deletions. Since the vertex minor is an NP-complete problem, we do not have an efficient algorithm and hence the algorithm we undertake, described below, has runtime that grows exponentially in $N$, the number of nodes in the larger of the two graphs presented to the classifier. 

Let $G$ and $G^\prime$ be the two graphs presented. Without loss of generality, let us assume $G$ has the larger (or equal) number of nodes of the two, which we will call the {\em parent graph}, and $G'$ the {\em child graph}. We then calculate $z$, the difference in the number of nodes between the two graphs. We then make $G$ go through all possible sequences of local complementations (LCs). A crucial point to note here is that in local complementation, we do not repeat vertices since $\tau_{v}\circ {\tau_{v}}(G) = G$. Hence, the total number of such sequences is $2^{N}$. For the graph produced by each LC sequence, we delete all possible combinations of $z$ nodes. The total number of ways deleting $z$ distinct vertices (without paying heed to any graph symmetries) is $2^z$. Finally, we check if the graph produced for each LC sequence and each vertex deletion set, is isomorphic to $G'$ or not.  The time complexity of this algorithm is:
 \begin{equation}\label{eq:37}
 O(N \times 2^{2N} \times 2^{-N^\prime}).
 \end{equation}
 where $N^\prime = |G^\prime|$ is the number of nodes in $G'$. If $N^\prime=N$, there is a polynomial-time algorithm which checks if $G$ and $G^\prime$ are vertex minors or not. This is because the problem of deciding if two graphs are LC-equivalent admits a polynomial-time algorithm~\cite{van2004efficient}. Eq.~\ref{eq:37} clearly shows that our algorithm is suboptimal since its time scaling does not become poly($N$) when $N^\prime=N$. But, since we used the brute force algorithm only to generate our training data (once) for $N$ ranging from $6$ to $12$, we did not look for a more efficient algorithm.

\section{Loss function derivation}\label{appendix:loss}

Here we focus on deriving the loss function for our GBS that undergoes photon loss from coupling efficiency, on-chip losses and detector efficiency. Losses can be added to the device with the help of a vacuum state coupled with the help of a  beamsplitter whose transmissivity is the transmissivity of the photon loss. It can be proven that if there is an equal loss on the output modes of an MZ interferometer, it is equivalent to having the same equal loss in the input modes as shown in Fig.\ref{fig:loss}(a).
With the help of this, we can push back all the loss terms from the on-chip and detector efficiency to the input modes and get a system where all the losses are in the input as shown in Fig.\ref{fig:loss}(b).

\section{Randomized Classical Algorithm Derivation}\label{appendix:random}
Given $G_1$ and $G_2$, where $G_2$ is a smaller graph and not a vertex minor of $G_1$. Let $G_3$ be a graph which is locally equivalent to $G_1$, that is, we obtained $G_3$ by a series of local complementations on $G_1$ denoted as:-

\begin{equation}\label{eq:38}
    G_3= {\tau_{v_i}}\circ{{\tau}_{v_{i-1}}} \ldots \circ{\tau_{v_1}}(G_1),
\end{equation}
where $i$ is an arbitrary integer.
We prove that “$G_2$ is not a vertex-minor of $G_3$” iff “$G_2$ is not a vertex-minor of $G_1$”. To prove this, we need to prove the following claims:-
\subsection{If $G_2$ is not a vertex-minor of $G_3$ then $G_2$ is not a vertex-minor of $G_1$}
We use proof by contradiction to prove this. Let $G_2$ be a vertex-minor of $G_1$. Then, 
\begin{equation}\label{eq:39}
    G_2= {\chi_{u_k}\circ \chi_{u_{k-1}} \ldots \circ \chi_{u_1}}\circ{\tau_{v_j}}\circ{{\tau}_{v_{j-1}}} \ldots \circ{\tau_{v_1}}(G_1),
\end{equation}
where $k$ and $j$ are arbitrary integers.
Since local complementation is reversible,from Eq.\ref{eq:38}, we can derive that,
\begin{equation}\label{eq:40}
    G_1= {\tau_{v_1}}\circ{{\tau}_{v_{2}}} \ldots \circ{\tau_{v_i}}(G_3),
\end{equation}
Substituting Eq.\ref{eq:40} in Eq.\ref{eq:39}, we get,
\begin{align}
\begin{split}\label{eq:41}
    G_2= {\chi_{u_k}\circ \chi_{u_{k-1}} \ldots \circ \chi_{u_1}}\circ{\tau_{v_j}}\circ{{\tau}_{v_{j-1}}} \ldots \circ{\tau_{v_1}}\\
    \circ{\tau_{v_1}}\circ{{\tau}_{v_{2}}} \ldots \circ{\tau_{v_i}}(G_3),
\end{split}
\end{align}

Hence from Eq.\ref{eq:41}, we can conclude that $G_2$ is a vertex-minor of $G_3$. But this is a contradiction. Hence, $G_2$ can never be a vertex minor of $G_1$. 
\subsection{If $G_2$ is not a vertex-minor of $G_1$ then $G_2$ is not a vertex-minor of $G_3$}
We use proof by contradiction to prove this. Let $G_2$ be a vertex-minor of $G_3$. Then, 
\begin{equation}\label{eq:42}
    G_2= {\chi_{u_k}\circ \chi_{u_{k-1}} \ldots \circ \chi_{u_1}}\circ{\tau_{v_j}}\circ{{\tau}_{v_{j-1}}} \ldots \circ{\tau_{v_1}}(G_3),
\end{equation}
Substituting Eq.\ref{eq:38} in Eq.\ref{eq:42}, we get:-
\begin{align}
\begin{split}\label{eq:43}
    G_2= {\chi_{u_k}\circ \chi_{u_{k-1}} \ldots \circ \chi_{u_1}}\circ{\tau_{v_j}}\circ{{\tau}_{v_{j-1}}} \ldots \circ{\tau_{v_1}}\\
    \circ{\tau_{v_i}}\circ{{\tau}_{v_{i-1}}} \ldots \circ{\tau_{v_1}}(G_1),
\end{split}
\end{align}
Hence from Eq.\ref{eq:43}, we can conclude that $G_2$ is a vertex-minor of $G_1$. But this is a contradiction.
Hence, $G_2$ can never be a vertex minor of $G_3$.
\\
Hence we can conclude that “$G_2$ is not a vertex-minor of $G_3$” iff “$G_2$ is not a vertex-minor of $G_1$”.

\section{Time complexity of GBS algorithm}\label{appendix:complexity}

 In this Appendix, we will calculate the time complexity of the GBS algorithm, both for when it is realized on a classical computer, as well as when it is executed on a real (quantum) GBS device.
 
 \subsection{Classical realisation of the GBS algorithm} 
 
 The first step of our quantum algorithm is to embed our two graphs in the GBS and generate samples. We convert the adjacency matrix of the 2 graphs to its respective covariance matrix using the Eq \eqref{eq:2}. This has a time complexity of $O(N^3)$. The sampling algorithm we use has a time complexity of $O(mN^32^{N/2})$ where $m$ is the number of modes and $N$ is the number of photons which is proportional to the $N \times N$ matrix embedded in the GBS. For our device, the number of modes is equal to the number of vertices in the graph, i.e., $m=N$. Hence the time complexity of the sampling algorithm becomes $O(N^42^{N/2})$ \cite{quesada2022quadratic}.  We then concatenate these two samples and pass them through the linear SVM with a time complexity of $O(N)$. This is then followed by the method of repeated trials to increase our accuracy. The complexity of repeated trials has already been derived in the above section and we use $n_{\rm q}(N)$ as the corresponding length of repeated trials to get a 97$\%$ accuracy for the given $E_{\rm q}(N)$. Hence, our final time-complexity for our quantum algorithm-classical is given as:-
  \begin{align}\label{eq:44}
  \begin{split}
 T_\text{cgbs}&=O(N^3)+n_{\rm q}(N) \times
 \bigl( O(N^42^{N/2})+ O(N)\Bigr) \\
 &=O(N^3)+n_{\rm q}(N) \times
  O(N^42^{N/2})\\
   &=n_{\rm q}(N) \times
  O(N^42^{N/2}),
  \end{split}
   \end{align}
where the last equality in the above scaling argument assumes that $n_{\rm q}(N)$ is an increasing function of $N$.

\subsection{Quantum realization of the GBS algorithm}

The first step of our quantum algorithm is to embed our two graphs in the GBS and generate samples. Embedding the graph in our GBS can be done by doing the Takagi decomposition of the adjacency matrix to give us the squeezing parameters and the unitary. The time complexity for the Takagi decomposition is $O(N^{2})$ \cite{xu2009twisted}. Here, our sampling algorithm is described in the optical propagation section and hence it has a time complexity of $O(N)$ where N is the number of vertices in the graph. We then concatenate the two samples and pass them through the linear SVM which has a time complexity of $O(N)$. This is then followed by repeated trials to increase our accuracy and we use $n_{\rm q}(N)$ as the corresponding length of repeated trials to get a 97$\%$ accuracy for the given $E_{\rm q}(N)$. Let the time taken to do a Takagi-Autonne decomposition for an $N$-node graph be $ t_{\rm{AT}}$ and the time taken for SVM to classify an input quantum feature vector of length $2N$ be $ t_{\rm{SVM}}$. Assuming that at time $t=0$ the first squeezed-light pulse is sent to the GBS, the total time taken for our algorithm after all the runs is equal to the time taken for our last sample to be classified by our SVM which is:-
\begin{equation}
    \label{45}
    t_\text{qgbs}=  t_{\rm{AT}} + ((n_{\rm q}(N)-1) \times  t_{\rm r}) +  t_{\rm o} +  t_{\rm{SVM}}
\end{equation}
    
Since $t_{\rm r}$ is a constant and $t_{\rm o}$ has an upper bound time complexity of $O(N)$, our final time-complexity for our quantum algorithm-GBS is given as:-
  \begin{align}
  \label{eq:46}
  \begin{split}
  T_\text{qgbs}&=O(N^2) + n_{\rm q}(N)  + O(N)\\
  &=O(N^2) + n_{\rm q}(N).
  \end{split}
\end{align}

\section{Encoding the graph in the GBS}\label{appendix:encode}

We see that if we interpret $\sigma$ as the covariance matrix of the output of an $N$-mode GBS, the GBS sample $\textbf{n}$ is drawn from a p.m.f. that is proportional to $\textnormal{Haf}^2(A_{\textbf{n}})$ per Eq.~\eqref{eq:3}, which in turn is a function $\textbf{n}$ and the adjacency matrix $A$ of the graph $G$ encoded into the GBS.

Any pure $N$-mode Gaussian state (described by a covariance matrix $\sigma$) can be obtained by a GBS, i.e., a linear-optical (passive) $N$-mode unitary $U$ acting on the $N$-mode tensor-product quantum state $|0;r_1\rangle |0;r_2\rangle \ldots |0;r_N\rangle$, where $|0;r_i\rangle$ is a squeezed vacuum state with squeezing parameter $r_i \ge 0$. Squeezing measured in dB (decibels) is $10\log_{10}(e^{2r})$. To obtain $U$, and $r_1, \ldots, r_N$, that would result in the state $\sigma$, one uses Takagi {\em et al.}'s decomposition~\cite{autonne1915matrices} of $A$:
\begin{equation}\label{eq:47}
    A= U\textnormal{diag}(\lambda_1,....,\lambda_M)U^T,
\end{equation}
where the squeezing parameters are given by $r_i= \textnormal{tanh}^{-1}(c\lambda_i), 1 \le i \le N$. The $U$ is the unitary that characterizes the interferometer, which in turn can be used to calculate the $N(N-1)$ single-mode phases in a universal programmable linear optical interferometer comprising of $N(N-1)/2$ Mach Zehnzer Interferometers (MZIs)~\cite{clements2016optimal}. Setting the scaling constant $c$ close to $1/s_{\rm{max}}$ could result in unrealistically-high single-mode squeezing values. One can clamp the maximum per-mode squeezing to $x$ dB, i.e., a maximum squeezing parameter $r$ per mode, with $x= 10\log _{10}(e^{2r})$, by picking
\begin{equation}\label{eq:48}
   c= \dfrac{\textnormal{tanh}(r)}{\lambda_{\rm{max}}},
\end{equation}
\\
where $\lambda_{\rm{max}} = \max(\lambda_1, \ldots, \lambda_N)$.

\section{Overview of the classical computer features}\label{appendix:features}
\begin{tabular}{ |p{3cm}|p{5cm}|  }
\hline
\multicolumn{2}{|c|}{Macbook Pro (15-inch, 2017)} \\
\hline
Features& Description  \\
\hline
Processor & 3.1 GHz Quad-Core Intel Core i7  \\
Memory & 16 GB 2133 MHz LPDDR3    \\
Graphics &Intel HD Graphics 630 1536 MB  \\
Version    &12.6.5   \\
Python version & 3.8.3  \\
Source-code editor & VSCode \\
\hline
\end{tabular}
\end{appendices}
\IEEEtriggeratref{4}
\end{document}